\documentclass[11pt,a4paper]{article}
\usepackage{graphicx}
\usepackage{epsf,epsfig,amssymb}

\def\bi{{\bf i}}
\def\bj{{\bf j}}

\def\bq{{\bf q}}
\def\bk{{\bf k}}

\def\cV{{\cal V}}

\def\Im{\mathrm{Im}}
\def\Re{\mathrm{Re}}

\def\up{\uparrow}
\def\down{\downarrow}

\def\eps{\epsilon}

\def\Gam{\Gamma}
\def\Lam{\Lambda}
\def\om{\omega}
\def\sg{\sigma}
\def\Sg{\Sigma}

\begin{document}




\title{Magnetic and superconducting correlations in the 
 two-dimensional Hubbard model}


\author{W. Metzner} 
\author{J. Reiss}
\author{{\bf W. Metzner, J. Reiss, D. Rohe}\\Max Planck Institute for Solid State Research\\
 Heisenbergstr. 1, D-70569 Stuttgart, Germany}
\maketitle  

\begin{abstract}
The interplay and competition of magnetic and superconducting
correlations in the weakly interacting two-dimensional Hubbard Model 
is investigated by means of the functional renormalization group. 
At zero temperature the flow of interactions in one-loop approximation
evolves into a strong coupling regime at low energy scales, signalling
the possible onset of spontaneous symmetry breaking. 
This is further analyzed by a mean-field treatment of the strong
renormalized interactions which takes into account magnetic and 
superconducting order simultaneously. 
The effect of strong correlations on single-particle properties in 
the normal phase is studied by calculating the flow of the self-energy.
\end{abstract}






\section{Introduction}

The two-dimensional Hubbard model has been proposed by Anderson
\cite{And} as a basic model for the electronic degrees of freedom in 
the copper-oxide planes of high-temperature superconductors.
In agreement with the generic phase diagram of the cuprates, the
Hubbard model exhibits antiferromagnetic order at half-filling, and 
is expected to become a d-wave superconductor near half-filling in 
two dimensions \cite{Sca}.
The exchange of antiferromagnetic spin fluctuations has been 
suggested as a plausible mechanism leading to d-wave pairing 
\cite{MSV,SLH,BSS}. In this picture the BCS 
effective interaction $V_{\bk\bk'}$ is roughly proportional to the 
spin correlation function $\chi_s(\bk-\bk')$. Close to
half-filling, $\chi_s(\bq)$ has a pronounced maximum at the 
antiferromagnetic wave vector $(\pi,\pi)$. As a consequence,
the gap equation with $V_{\bk\bk'}$ as input has a solution with
$d_{x^2-y^2}$-wave symmetry, such that the gap $\Delta_{\bk}$
has maximal modulus but opposite sign near the points $(\pi,0)$ 
and $(0,\pi)$ in the Brillouin zone, respectively. 
This intuitive argument has been corroborated by a self-consistent
perturbative solution of the two-dimensional Hubbard model within 
the so-called fluctuation-exchange approximation \cite{BSW}.

The spin-fluctuation mechanism for pairing could in principle be 
spoiled by other contributions to the BCS interactions and also by 
spin density wave instabilities. Unfortunately,
it is hard to prove the existence of superconductivity in the 
Hubbard model by exact numerical computation \cite{Sca,Dag}, 
as a consequence of finite size and/or temperature limitations.
However, the tendency towards antiferromagnetism and d-wave 
pairing is captured already by the 2D Hubbard model at {\em weak}\/
coupling, where one can hope to proceed by perturbative
calculations. 
Conventional perturbation theory breaks down for densities close 
to half-filling, since competing infrared divergences appear as a 
consequence of Fermi surface nesting and van Hove singularities 
\cite{Sch87,Dzy87,LMP}.
A controlled and unbiased treatment of these divergencies cannot
be achieved by standard resummations of Feynman diagrams, but
requires a renormalization group (RG) analysis which takes into 
account the particle-particle and particle-hole channels on an equal 
footing. 
In the following we describe progress in this direction achieved 
by applying the socalled \emph{functional renormalization group} 
(fRG) to the two-dimensional Hubbard model.

In Sec.\ 2 we define the two-dimensional Hubbard model and describe
the relevant infrared singularities in perturbation theory.
In Sec.\ 3 we briefly review the Wick ordered version of the functional 
renormalization group and discuss the one-loop flow of effective 
interactions and susceptibilities for the Hubbard model.
In Sec.\ 4 we compute the flow of the electronic self-energy and
discuss how the strong correlations at low energy scales affect
spectral properties of single-particle excitations. 
In Sec.\ 5 we treat the effective low-energy problem obtained from the
fRG flow at zero temperature within a mean-field approximation and find 
a range of fillings where magnetic and and sizable superconducting 
order coexist. 
A short conclusion follows in Sec.\ 6.

\section{Two-dimensional Hubbard Model}

The Hubbard model describes tight-binding electrons with a local 
repulsion $U>0$. The Hamiltonian reads
\begin{equation}
 H = \sum_{\bi,\bj} \sum_{\sg} t_{\bi\bj} \,
 c^{\dag}_{\bi\sg} c_{\bj\sg} +
 U \sum_{\bj} n_{\bj\up} n_{\bj\down} \; ,
\end{equation}
where $c^{\dag}_{\bi\sg}$ and $c_{\bi\sg}$ are creation and 
annihilation operators for fermions with spin projection 
$\sg = \up,\down$ on a lattice site ${\bi}$,
and $n_{\bj\sg} = c^{\dag}_{\bj\sg} c_{\bj\sg}$.
A hopping amplitude $-t$ between nearest neighbors and
an amplitude $-t'$ between next-nearest neighbors on a square
lattice leads to the dispersion relation
\begin{equation}
 \eps_{\bk} = -2t(\cos k_x + \cos k_y)
              -4t'(\cos k_x \, \cos k_y)
\end{equation}
for single-particle states.
This dispersion relation has saddle points at $\bk = (0,\pi)$
and $(\pi,0)$, which lead to logarithmic van Hove singularities 
in the non-interacting density of states at the energy 
$\eps_{\rm vH} = 4t'$.

For a chemical potential $\mu = \eps_{\rm vH}$ the Fermi 
surface contains the van Hove points.
In this case a perturbative calculation of the two-particle 
vertex function leads to several infrared divergencies already 
at second order in $U$, that is at one-loop level 
\cite{Sch87,Dzy87,LMP}.
In particular, the particle-particle channel diverges as $\log^2$ 
for vanishing total momentum $\bk_1+\bk_2$, and logarithmically for 
$\bk_1+\bk_2 = (\pi,\pi)$.
The particle-hole channel diverges logarithmically for vanishing
momentum transfer; for momentum transfer $(\pi,\pi)$ it diverges
logarithmically if $t' \neq 0$ and as $\log^2$ in the special 
case $t'=0$.
Note that $\mu=0$ for $t'=0$ corresponds to half-filling ($n=1$).
In this case there are also logarithmic divergences 
for all momentum transfers parallel to $(\pi,\pi)$ or $(\pi,-\pi)$
due to the strong nesting of the square shaped Fermi surface.
For $\mu \neq \eps_{\rm vH}$ only the usual logarithmic Cooper 
singularity at zero total momentum in the particle-particle 
channel remains.
However, the additional singularities at $\mu = \eps_{\rm vH}$ 
lead clearly to largely enhanced contributions for small 
$|\mu - \eps_{\rm vH}|$, especially if $t'$ is also 
small.

Hence, for small $|\mu - \eps_{\rm vH}|$ one has to deal with 
competing divergencies in different channels.
This problem calls for a renormalization group treatment.

\section{Renormalization Group Approach}

Early RG studies of the two-dimensional Hubbard model started with 
simple scaling approaches, very shortly after the discovery of 
high-$T_c$ superconductivity \cite{Sch87,Dzy87,LMP}. 
These studies focused on dominant scattering processes between 
van Hove points in k-space, for which a small number of running 
couplings could be defined and computed on one-loop level. 
Spin-density and superconducting instabilities were identified
from divergencies of the corresponding correlation functions.

A complete treatment of all scattering processes in the Brillouin
zone is rather complicated since the effective interactions cannot 
be parametrized accurately by a small number of variables,
even if irrelevant momentum and energy dependences are neglected.
The {\em tangential}\/ momentum dependence of effective interactions 
along the Fermi surface is strong and important in the low-energy 
limit. Hence, one has to deal with the renormalization of coupling
{\em functions}.
This problem is treated most naturally within a Wilsonian momentum 
shell RG \cite{WK}, where modes are integrated out successively 
from high energy states down to the Fermi surface \cite{Sha}.
This type of RG is also the basis for important rigorous work on
two-dimensional Fermi systems \cite{Sal}.
The successive integration of modes can be formulated as an
exact hierarchy of flow equations for effective interactions
(one-particle, two-particle etc.), which is known as the {\em exact} 
or {\em functional} renormalization group \cite{WH,Pol,Wet}.
Starting with the work of Zanchi and Schulz \cite{ZS}, several 
groups have computed effective interactions and susceptibilities
for the two-dimensional Hubbard model, using various versions of 
the fRG in one-loop approximation \cite{HM00a,HSFR}.
Here we focus on results obtained from the Wick-ordered version 
of the fRG \cite{Sal}.
Other versions yield qualitatively similar results.

\subsection{Wick ordered flow equations}

The renormalization group equations are obtained as follows
(for details, see Salmhofer \cite{Sal} and Ref.\ \cite{HM00a}).
The infrared singularities are regularized by introducing an
infrared cutoff $\Lam > 0$ into the bare propagator such that
contributions from momenta with $|\eps_{\bk} - \mu| < \Lam$ 
are suppressed.
All Green functions of the interacting system will then depend 
on $\Lam$, and the true theory is recovered only in the limit 
$\Lam \to 0$.
The RG equations are most conveniently obtained from the
effective potential $\cV^{\Lam}$, which is the 
generating functional for connected Green functions with bare 
propagators amputated from the external legs.
Taking a $\Lam$-derivative one obtains an exact functional flow
equation for this quantity.
Expanding $\cV^{\Lam}$ on both sides of the flow equation in 
powers of the fermionic fields (i.e.\ Grassmann variables),
and comparing coefficients, one obtains the so-called Polchinski 
equations \cite{Pol} for the effective m-body interactions used by 
Zanchi and Schulz \cite{ZS}.
Salmhofer \cite{Sal} has pointed out that an alternative 
expansion in terms of {\em Wick ordered}\/ monomials
of fermion fields
yields flow equations for the corresponding m-body interactions 
$V^{\Lam}_m$ with a particularly simple structure (see Fig.\ 1).
\begin{figure}[ht]
\centering
\includegraphics[width = 10cm]{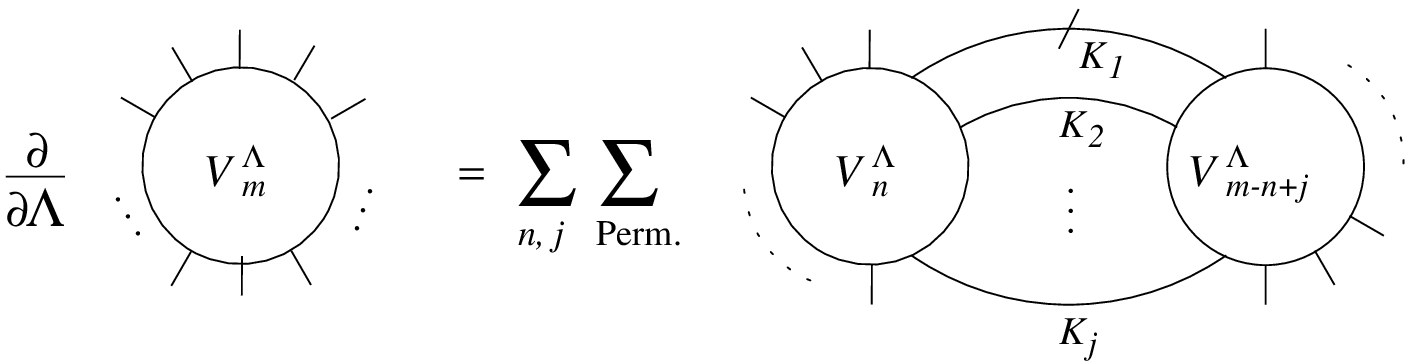}
\caption{Diagrammatic representation of the flow equation
 for $V^{\Lam}_m$ in the Wick ordered version of the fRG. 
 The line with a slash corresponds to 
 $\partial D^{\Lam}/\partial\Lam$,
 the others to $D^{\Lam}$; all possible pairings leaving $m$ 
 incoming and $m$ outgoing external legs have to be summed.}
\label{fig1}
\end{figure}
The flow of $V^{\Lam}_m$ is given as a bilinear form of other
n-body interactions (at the same scale $\Lam$), which are 
connected by lines corresponding to the propagator
\begin{equation}
 D^{\Lam}(k) = \frac{\Theta(\Lam - |\xi_{\bk}|)}{ik_0 - \xi_{\bk}}
 \; ,
\end{equation}
where $\xi_{\bk} = \eps_{\bk} - \mu$, and one line corresponding 
to $\partial D^{\Lam}(k)/\partial\Lam$.
For small $\Lam$, the momentum integrals on the right hand side
of the flow equation are thus restricted to momenta close to the
Fermi surface (see Fig.\ 2).
\begin{figure}[ht]
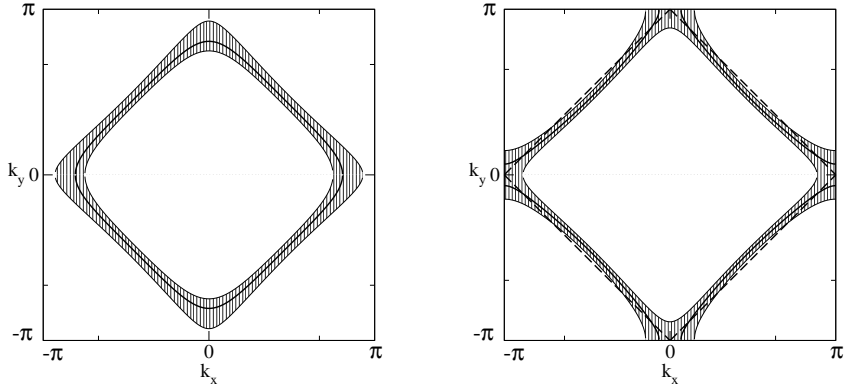

\centering
\vskip 7mm
\includegraphics[width = 5cm]{fig2a.eps}
\hskip 1.0cm
\includegraphics[width = 5cm]{fig2b.eps}
\caption{Fermi surfaces (solid lines) and support of the propagator 
 $D^{\Lam}$ (shaded regions) in momentum space; 
 {\em left}: $t'=0$, {\em right}: $t'>0$. The Fermi surface on the right
 intersects the socalled umklapp surface (dashed line).}
\label{fig2}
\end{figure}
With the initial condition $\cV^{\Lam_0} = \mbox{bare interaction}$,
where $\Lam_0 = \max |\xi_{\bk}|$, the above flow equations 
determine the {\em exact}\/ flow of the effective interactions as 
$\Lam$ sweeps over the entire energy range from the band edges down 
to the Fermi surface.
The {\em effective}\/ low-energy theory can thus be computed directly
from the {\em microscopic} theory without introducing any ad hoc
parameters.
The structure of these flow equations is very convenient for a 
power counting analysis to arbitrary loop order \cite{Sal},
and also for a concrete numerical solution.

\subsection{One-loop flow}

To detect instabilities of the system in the weak-coupling limit,
it is sufficient to truncate the infinite hierarchy of flow 
equations described by Fig.\ 1 at {\em one-loop}\/ level, 
and neglect all components of the effective interaction except the
two-particle interaction $V^{\Lam}_2$. 
The effective two-particle interaction $V^{\Lam}_2$ then reduces to 
the one-particle irreducible two-particle vertex $\Gam^{\Lam}$, and 
its flow is determined by $\Gam^{\Lam}$ itself (no other m-body 
interactions enter).
Putting arrows on the lines to distinguish creation and 
annihilation operators one thus obtains the flow equation
for $\Gam^{\Lam}$ shown graphically in Fig.\ 3.
\begin{figure}[ht]
\centering
\includegraphics[width = 10cm]{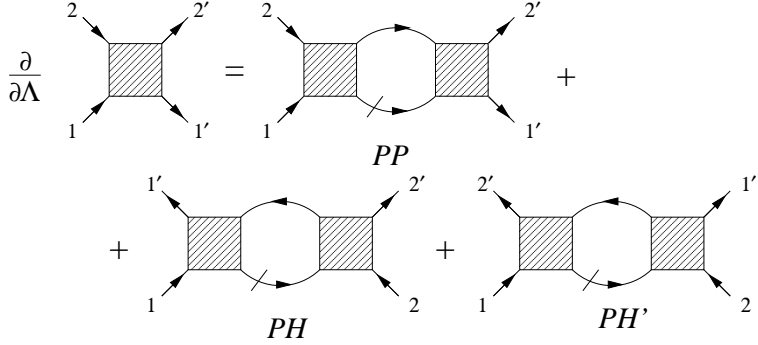}
\caption{Flow equation for the two-particle vertex
 $\Gam^{\Lam}$ in one-loop approximation with the particle-particle
 channel (PP) and the two particle-hole channels (PH and PH').}
\label{fig3}
\end{figure}
Flow equations for {\em susceptibilities} are obtained by considering
the exact fRG equations in the presence of suitable external 
fields, which leads to an additional one-body term in the bare
interaction, and expanding everything in powers of the external
fields to sufficiently high order \cite{HM00a}.
On one-loop level one obtains the flow equations shown in Fig.\ 4.
\begin{figure}[ht]
\centering
\includegraphics[width = 7cm]{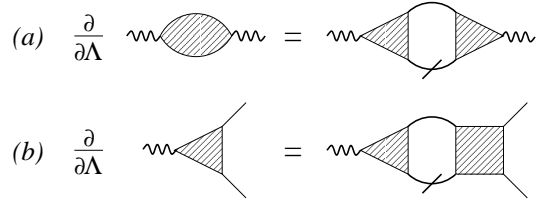}
\caption{Flow equations for (a) the susceptibilities
 $\chi^{\Lam}$ 
 and (b) the response vertices $R^{\Lam}$ in one-loop 
 approximation.}
\label{fig4}
\end{figure}
The flow of a susceptibility $\chi^{\Lam}$ is determined by
the corresponding response vertex $R^{\Lam}$, the flow of
which is in turn given by $\Gam^{\Lam}$ and $R^{\Lam}$ itself.
The initial conditions are given by $\chi^{\Lam_0} = 0$ for the
susceptibilities and by the bare response vertices for 
$R^{\Lam}$.
For pairing susceptibilities only the particle-particle
channel contributes to the propagator pair in Fig.\ 4, for
charge and spin density susceptibilities only the particle-hole
channel.

It is clearly impossible to solve the flow equations with the full
energy and momentum dependence of the two-particle vertex, since
$\Gam^{\Lam}$ has three independent energy and momentum variables.
The problem can however be simplified by ignoring dependences 
which are {\em irrelevant} (in the RG sense) in the low energy limit, 
namely the energy dependence and the momentum dependence normal to 
the Fermi surface.
Hence, we compute the flow of the two-particle vertex at zero energy
and with at least three momenta on the Fermi surface (the fourth
being determined by momentum conservation). On the right hand side
of the flow equation we approximate the vertex by its zero energy 
value with three momenta projected on the Fermi surface
(if not already there), as indicated in Fig.\ 5.
\begin{figure}[ht]
\centering
\includegraphics[width = 5cm]{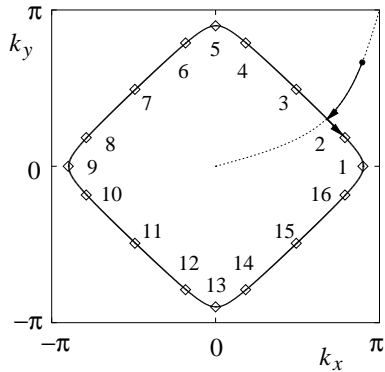}
\caption{Projection of momenta on the Fermi surface;
 discretization and labelling of angle variables.}
\label{fig5}
\end{figure}
This projection procedure is exact for the bare Hubbard vertex,
and asymptotically exact in the low-energy regime, since only
irrelevant dependences are neglected.
The remaining tangential momentum dependence is discretized.
The momentum-dependence of the two-particle vertex is thus 
approximated by a step function which is constant on ''patches'' 
(sectors) in the Brillouin zone.

The flow of the two-particle vertex has been computed for many 
different model parameters $t'$ and $U$ ($t$ just fixes the
absolute energy scale) and densities close to half-filling.
At $T=0$ the vertex develops a strong momentum
dependence for small $\Lam$ with divergencies for several
momenta at some critical scale $\Lam_c > 0$, which vanishes
exponentially for $U \to 0$.
To see which physical instability is associated with the
diverging two-particle vertex, various susceptibilities have
been computed, in particular:
commensurate and incommensurate spin susceptibilities 
$\chi_S(\bq)$ with $\bq = (\pi,\pi)$, $\bq = (\pi-\delta,\pi)$ and 
$\bq = (1-\delta) (\pi,\pi)$, where $\delta$ is a function 
of density \cite{Sch90}, the commensurate charge susceptibility 
$\chi_C((\pi,\pi))$, and singlet pair susceptibilities with form 
factors
\begin{equation}
 d(\bk) =  \left\{ \begin{array}{ll}
 1 & 
 \mbox{($s$-wave)} \\     
 \frac{1}{\sqrt{2}} (\cos k_x + \cos k_y) &
 \mbox{(extended $s$-wave)} \\
 \frac{1}{\sqrt{2}} (\cos k_x - \cos k_y) &
 \mbox{($d$-wave $d_{x^2-y^2}$)} \\
 \sin k_x \sin k_y &
 \mbox{($d$-wave $d_{xy}$)}. \end{array} \right. \; .
\end{equation}
Some of these susceptibilities diverge together with the 
two-particle vertex at the scale $\Lam_c$.
Depending on the choice of $U$, $t'$ and $\mu$ the strongest
divergence is found for the commensurate or incommensurate
spin susceptibility or for the pair susceptibility with 
$d_{x^2-y^2}$ symmetry. 
In Fig.\ 6 we show a typical result for the flow of the 
two-particle vertex and susceptibilities as a function of $\Lam$.
\begin{figure}[ht]
\centering
\includegraphics[width = 5.8cm]{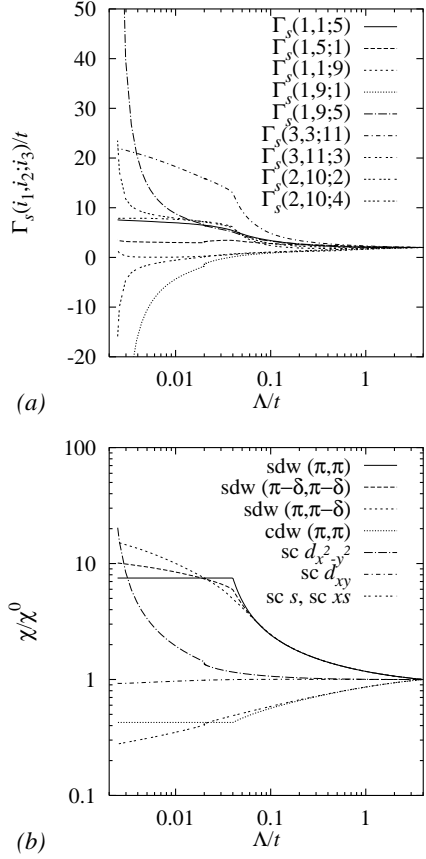}
\caption{(a) The flow of the singlet vertex 
 $\Gam^{\Lam}_s$ as a function of $\Lam$ for several choices of 
 the momenta $\bk_{F1}$, $\bk_{F2}$ and $\bk'_{F1}$, which are
 labelled according to the numbers in Fig.\ 5. 
 The model parameters are $U = t$ and $t'= 0$, the chemical 
 potential is $\mu = - 0.02 t$;
 (b) the flow of the ratio of interacting and non-interacting 
 susceptibilities, $\chi^{\Lam}/\chi_0^{\Lam}$, for the same 
 system [from Ref.~\cite{HM00a}].}
\label{fig6}
\end{figure}
Only the singlet part (from a spin singlet-triplet decomposition) 
of the vertex is plotted, for various choices of the three independent 
external momenta.
The triplet part flows generally more weakly than the singlet part. 
Note the threshold at $\Lam=2|\mu|$ below which the amplitudes for 
various scattering processes, especially umklapp scattering, 
renormalize only very slowly. The flow of the antiferromagnetic 
spin susceptibility is cut off at the same scale.
The infinite slope singularity in some of the flow curves at
scale $\Lam=|\mu|$ is due to the van Hove singularity being 
crossed at that scale.
The {\em pairing susceptibility}\/ with $d_{x^2-y^2}$-symmetry is
obviously {\em dominant}\/ here (note the logarithmic scale).
Following the flow of the two-particle vertex and susceptibilities,
one can see that those interaction processes which enhance
the antiferromagnetic spin susceptibility (especially umklapp
scattering) also build up an attractive interaction in the 
$d_{x^2-y^2}$ pairing channel. This confirms the spin-fluctuation 
route to $d$-wave superconductivity \cite{Sca}.
Similar results leading to the same conclusions have been obtained
by other one-loop calculations using different versions of the fRG
\cite{ZS,HSFR,KK03a}.

\section{Two-loop self-energy}

The self-energy $\Sg$ for the 2D Hubbard model, which encodes the 
influence of correlations on single-particle excitations, has been 
analyzed within the functional RG framework in several recent 
articles. 
Interesting dynamical contributions to the self-energy set in at 
second order in the two-particle interactions, corresponding to 
the two-loop diagram shown in Fig.~7. 
\begin{figure}[ht]
\centerline{\includegraphics[width = 5cm]{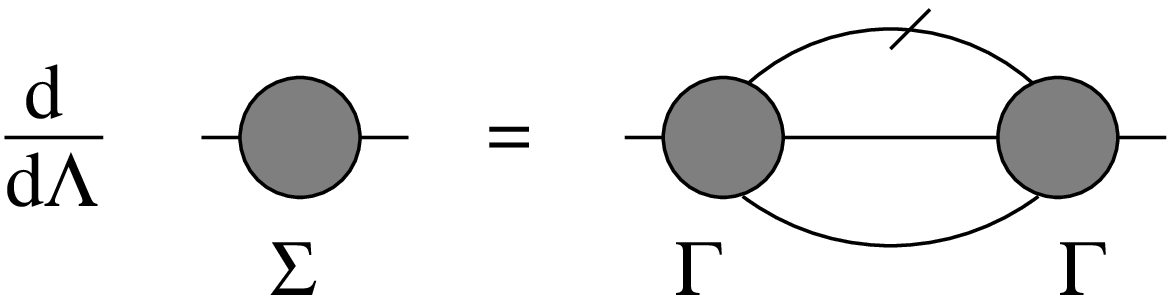}}
\caption{Flow equation for the two-loop self energy.}
\label{fig7}
\end{figure}
The vertices on the right-hand side were obtained from the one-loop
flow equation. A full two-loop calculation, including two-loop 
vertex renormalization, would be very complex, and has not been 
done so far. 

Some of the earlier studies focused on the frequency derivative of
the self-energy on the Fermi surface, which determines the wave 
function renormalization factor $Z = 
[1 - \left. \partial_{\om}\Re\Sg(\bk_F,\om) \right|_{\om=0}]^{-1}$.
In Fermi liquids $Z$ yields the weight of the quasi-particle peak
in the spectral function.
Zanchi \cite{Zan} computed the flow of $Z$ for the 2D Hubbard model
with pure nearest neighbor hopping at half-filling and found that
$Z$ is strongly reduced when the renormalized interactions increase,
the suppression being strongest near the van Hove points.
Subsequently, Honerkamp and Salmhofer \cite{HS} computed the flow
of the $Z$-factor at and also away from half-filling. 
While confirming an anisotropic reduction of $Z$, they observed
that this suppression evolves much slower than the divergence of 
dominant interactions. In addition, they pointed out that the 
self-energy does not significantly affect the vertex renormalization
in the perturbative regime, that is before the interactions become
very strong. 
The imaginary part of the self-energy at zero frequency, 
$\Im\Sg(\bk_F,0)$, was computed by Honerkamp \cite{Hon}. 
In Fermi liquids this quantity is directly related to the decay rate 
of quasi-particles.
For a hole-doped Hubbard model with a concave Fermi surface that 
almost touches the van Hove points (this requires a negative $t'$)
he found that $\Im\Sg(\bk_F,0)$ is moderately anisotropic, with larger 
values for momenta closer to the van Hove points. 

Functional RG calculations of the full frequency dependence of the 
self-energy were performed recently by Katanin and Kampf \cite{KK04}
within the one-particle irreducible version of the fRG, and by 
two of us \cite{RM} within the Wick ordered version.
The real and imaginary parts of $\Sg(\bk_F,\om)$ and the resulting
spectral function $A(\bk_F,\om)$ were computed for the 2D Hubbard 
model with nearest and next-to-nearest neighbor hopping at finite
temperature, which was chosen such that the largest renormalized 
interactions were strongly enhanced, but did not diverge for 
$\Lam \to 0$.
Marked deviations from Fermi liquid behavior were obtained for $\bk_F$ 
close to a van Hove point, and also near other hot spots \cite{hot}
in the case of filling factors above the van Hove singularity:
the imaginary part of $\Sg(\bk_F,\om)$ develops a pronounced peak
at low frequencies, as seen for example in Fig.~8,
which leads to a double-peak structure in the spectral functions,
reminiscent of a pseudogap. 
\begin{figure}[ht]
\centerline{\includegraphics[width = 9cm]{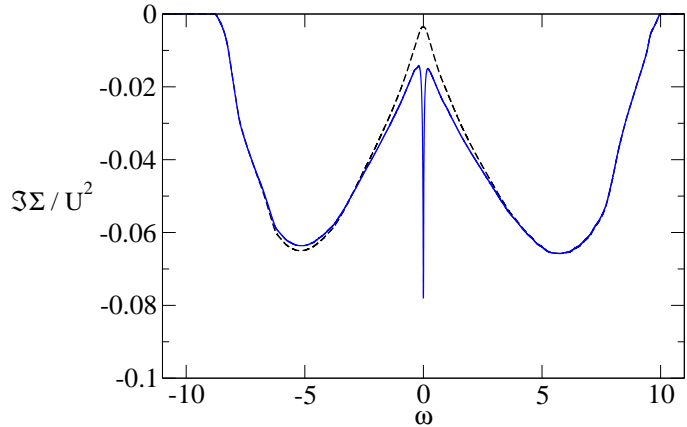}}
\caption{Imaginary part of the self energy $\Im\Sg(\bk_F,\om)$
 normalized by $U^2$ as a function of $\om$ and $\bk_F$ on a
 hot spot, for parameters $t'/t = -0.1$, $T/t = 0.05$,
 $U/t = 1.8$ and density $n = 0.92$; 
 the result from the functional RG (solid blue line) is
 compared to the result from second order perturbation theory
 (dashed black line) [from Ref.~\cite{RM}].}
\label{fig8}
\end{figure}
This double peak gradually transforms into a single peak when 
moving away from the hot spot or van Hove regions, and also upon
raising the temperature.
In the one-particle irreducible scheme the peak in 
$\Im\Sg(\bk_F,\om)$ was found to be partially split by a dip
upon moving away from the van Hove point, which leads to a three
peak structure in the spectral function \cite{KK04}.
No such splitting was obtained in the Wick ordered functional 
RG \cite{RM}. 
This discrepancy can be traced back to differences in the 
approximate flow equations of both schemes, and indicates
that higher order corrections become important when the 
renormalized interactions grow too strong. 

One can conclude that self-energy contributions are particularly
large near hot spots and tend to destroy Fermi liquid behavior,
but the final form of the spectral function requires an analysis
beyond the above perturbative truncation of the flow equations.

\section{Flow to strong coupling: RG + Mean-Field}

Within the one-loop truncation the renormalized two-particle 
interaction $\Gam^{\Lam}$ always diverges at $T=0$ in one or 
several momentum channels at a finite energy scale $\Lam_c$, 
even for a small bare interaction $U$, resulting in a strong 
coupling problem in the low-energy limit.
If the two-particle vertex diverges only in the Cooper channel,
driven by the particle-particle contribution to the flow,
this strong coupling problem can be treated by exploiting 
$\Lam_c$ as a small parameter \cite{FMRT}. 
The scale $\Lam_c$ is exponentially small for a small bare 
interaction. 
The formation of a superconducting ground state can then be 
described essentially by a BCS theory with renormalized input 
parameters. 
A method very similar in spirit has been applied very recently to 
superconductivity in quasi-one-dimensional systems \cite{FUSE}.

Here, we use an analogous approach to treat \emph{several} order 
parameters simultaneously at mean-field level. Near half filling, 
for $U = 2.5t$ and $t'=-0.2t$ the one-loop flow produces strong 
interactions in the Cooper channel as well as in the particle-hole 
channel, the latter being dominated by commensurate $(\pi,\pi)$ 
scattering. We thus concentrate on d-wave superconductivity and 
antiferromagnetism, which can be treated within a $4 \times 4$ 
Nambu formalism. While the interplay between antiferromagnetic 
order and d-wave superconductivity has been studied earlier within 
mean-field theories using models containing pairing interactions 
\cite{INU,KYU}, we emphasize that we do not introduce such couplings 
\emph{ad hoc}. Rather, pairing interactions arise within the 
purely repulsive Hubbard model via the fRG treatment.   

In an extended RG framework, spontaneous symmetry breaking can be 
handled by adding an infinitesimal symmetry breaking term at the 
beginning of the flow, which is then promoted to a finite order 
parameter at the scale $\Lam_c$ \cite{SHML}. 
However, this approach would be extremely involved for the case 
of competing order parameters we are interested in here. 
Instead, we stop the one-loop flow at a scale $\Lam_1$ that is 
small compared to the band width but still safely above the scale 
$\Lam_c$ where the two-particle vertex diverges. 
At this point the vertex has developed already a pronounced 
momentum dependence, reflecting in particular magnetic and 
superconducting correlations.
The integration over the remaining modes, below $\Lam_1$, is
treated in a mean-field approximation allowing antiferromagnetic
and superconducting order. The mean-field theory is
defined on a restricted momentum region near the Fermi surface, 
with $|\eps_{\bk}-\mu| < \Lam_1$, and the effective 
interactions entering the mean-field equations are extracted from 
$\Gam^{\Lam_1}$.
Details on this combined RG+MF approach will be presented 
elsewhere \cite{RRM05}.
We now discuss some results for the order parameters.

In Fig.\ 9 we show the amplitudes of the d-wave superconducting gap 
(SCG) and the s-wave antiferromagnetic order parameter (AFG) as a 
function of (electron) density. 
\begin{figure}[ht]
\centerline{\includegraphics[width = 8cm]{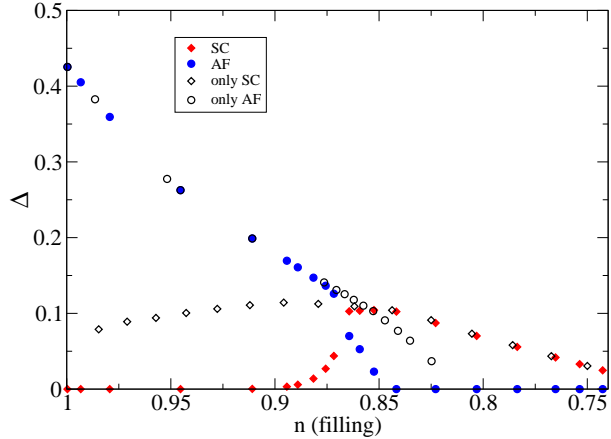}}
\caption{d-wave superconducting order parameter and antiferromagnetic 
 order parameter (in units of $t$) as a function of density for 
 $U/t = 2.5$ and $t'/t = -0.2$. 
 Filled colored symbols represent the results obtained for the 
 combined theory with two order parameters, while in the calculation 
 leading to the open symbols only one order parameter, either 
 antiferromagnetic or superconducting, was allowed in the mean-field
 equations.}
\label{fig9}
\end{figure}
The latter is computed simultaneously 
with the order parameters, and thus differs from the initial density 
of the bare system. For $n=1$, i.e. at half filling, the AFG is
largest, while the SCG is zero. In this case, the system is fully 
gapped. With decreasing filling the AFG decreases initially roughly 
linearly, while the SCG remains numerically zero. 
For electron densities below one, holes appear first in pockets 
around $(\pi/2,\pi/2)$, which define a surface of low-energy 
excitations of the non-half filled system which increases with 
larger hole doping. 
In principle, this residual Fermi surface is always unstable
against superconducitivity, due to attraction in the d-wave
Cooper channel.
However, very close to half-filling, where the pockets are small,
the SCG is tiny, since the d-wave attraction is very small near
the Brillouin zone diagonal. 
When the hole pockets are large enough, roughly at $n \approx 0.88$, 
a sizeable SCG develops, coexisting with a finite AFG. 
The numerical analysis is rather involved in the coexistence 
region, and the precise behavior of the order parameters in this
regime remains to be clarified.
What is clear is that there is a range of densities for which both 
order parameters are sizable.
The coexistence disappears below $n \approx 0.85$. Below this density 
the AFG truly vanishes, and the SCG is finite and decreases with 
decreasing filling. The respective results for both order parameters 
when the other one is set to zero are also shown. When the AFG is 
set to zero, the SCG persist even at half filling. When the SCG is 
set to zero, the AFG is enhanced in the coexistence region. In both 
cases, a finite value for one order parameter leads to a suppression 
of the other. 
While these results have been obtained for a weak on-site repulsion, 
they are in line with the behaviour at stronger coupling as found 
within variational cluster perturbation theory \cite{SLMT}. There, 
however, due to finite size effects the possibilty of coexistence 
on the hole doped side could not be conclusively answered. 

Fig.\ 10 shows the angular dependence of the SCG and the AFG for 
three selected densities, along with the corresponding hole pockets. 
Here, $\phi$ is the angle with respect to the $k_x$-axis.  
Results are shown for a discretisation of 48 patches. 
\begin{figure}[ht]
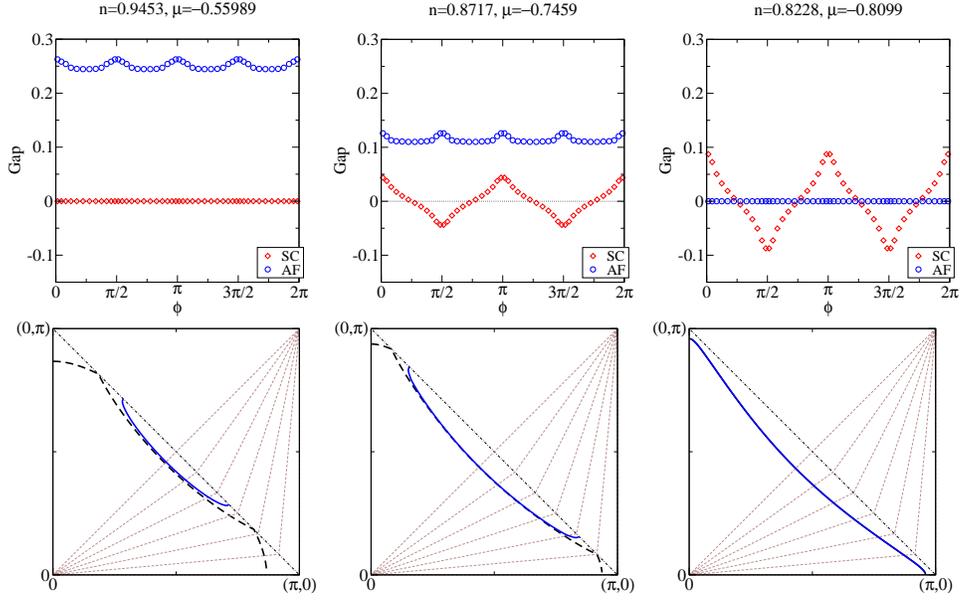

\includegraphics[angle=0,width=4.0cm]{fig10a.eps} 
\includegraphics[width=4.0cm]{fig10b.eps} 
\includegraphics[width=4.0cm]{fig10c.eps} \hfill
\phantom{.}\includegraphics[angle=0,width = 4.0cm]{fig10d.eps} 
\hspace{0.05cm}
\includegraphics[angle=0,width = 4.0cm]{fig10e.eps} \hspace{0.05cm}
\includegraphics[angle=0,width = 4.0cm]{fig10f.eps} \hfill
\caption{{\em Top}: d-wave superconducting and antiferromagnetic order 
 parameters at three different values of the chemical potential. 
 {\em Bottom}: Fermi surfaces in the magnetic Brillouin zone (solid blue 
 lines); in the antiferromagnetic state close to half-filling the 
 Fermi surface forms a hole pocket around $(\pi/2,\pi/2)$. 
 The corresponding bare Fermi surfaces, backfolded with respect 
 to Umklapp surface, are shown as broken lines. 
 The straight lines indicate the patching scheme.}
\label{fig10}
\end{figure}
For $n=0.9453$ we obtain a sizeable AFG which has s-wave symmetry and 
is slightly anisotropic. The SCG is practically zero. The oval 
pocket which opens around $(\pi/2,\pi/2)$ is sizeable, but does not 
extend far enough to allow for a sizeable d-wave SCG. 
For $n=0.8717$ we observe a coexistence of both order parameters. 
While the AFG is reduced in size compared to the case $n=0.9453$, 
its shape remains essentially the same. The pocket extends further 
away from the zone diagonal, and together with the smaller AFG this 
allows for a substantial d-wave SCG. In this situation the low-energy 
excitations are gapped due to the AFG near the points $(\pi,0)$ and 
$(0,\pi)$, and gapped due to the SCG along the ``Fermi pocket'', 
with nodes along the Brillouin zone diagonals.
For an even smaller density of $n=0.8228$ the AFG vanishes and only 
a d-wave SCG remains, which extends over the whole Fermi surface,
except at the nodal points.

While we have restricted our analysis to superconductivity and 
antiferromagnetism, there are other instabilities which may arise, 
such as ferromagnetism (at moderate $|t'/t|$ in the Hubbard model)
\cite{HSG,HS01,KK03b} or a d-wave Pomeranchuk instability of the 
Fermi surface \cite{HM00b,GKW}.
The treatment presented here neglects order parameter fluctuations, 
which are particularly important in low dimensions. 
Since we work at $T=0$ we expect the above results to be qualitatively 
stable.
In a more sophisticated approach order parameter fluctuations can be
treated most conveniently by introducing appropriate bosonic fields, 
as discussed recently for antiferromagnetic order in the half-filled 
Hubbard model at finite temperature \cite{BBW}. 

\section{Conclusion}

In summary, the functional renormalization group captures many 
aspects of the complex interplay of magnetic and superconducting
correlations in the two-dimensional Hubbard model.
In particular, it captures the generation of attractive d-wave
Cooper interactions via antiferromagnetic spin correlations
already in a one-loop approximation of the vertex flow. 
Strong correlation effects in the single particle excitations are 
obtained from the two-loop flow of the self-energy. Magnetic
fluctuations suppress spectral weight at hot spots, leading to 
a pseudogap shape of the spectral function at these points.
The strong-coupling problem posed by the divergence of the flow
at $T=0$ is treated approximately by combining the fRG with a 
mean-field treatment at low energies. Superconducting and 
antiferromagnetic instabilities are found to compete, with a 
small range of densities for which both orders can coexist
with a sizable order parameter for each.

A more complete analysis of the two-dimensional Hubbard model
by renormalization group methods, 
including also order parameter fluctuations at low energy scales,
can be expected to yield further important clues for a better 
understanding of magnetic and superconducting correlations in 
high temperature superconductors.

{\bf Acknowledgement}
We thank C.\ Honerkamp, A.A.\ Katanin, and M.\ Salmhofer 
for valuable discussions, and S.\ Andergassen for a critical
reading of the manuscript.


\begin{thebibliography}{10}
\bibitem{hot} Hot spots are intersections of the Fermi line and the
 magnetic Brillouin zone boundary.
\end{thebibliography}
\end{document}